# Geometry and neoclassical theory in a quasi-isodynamic stellarator

Matt Landreman, Peter J Catto

Massachusetts Institute of Technology, Plasma Science and Fusion Center,

167 Albany St., Cambridge, MA 02139, USA

E-mail: landrema@mit.edu

**Abstract**

We show that in perfectly quasi-isodynamic magnetic fields, which are generally non-quasisymmetric and which can approximate fields of experimental interest, neoclassical calculations can be carried out analytically more completely than in a general stellarator. Here, we define a quasi-isodynamic field to be one in which the longitudinal adiabatic invariant is a flux function and in which the constant-$B$ contours close poloidally. We first derive several geometric relations among the magnetic field components and the field strength. Using these relations, the forms of the flow and current are obtained for arbitrary collisionality. The flow, radial electric field, and bootstrap current are also determined explicitly for the long-mean-free-path regime.

November 23, 2010

# 1. Introduction

Drift trajectories of trapped particles in a stellarator are not generally confined, leading to large fluxes of particles and heat compared to a comparable axisymmetric plasma. It is therefore a goal of stellarator design to minimize the average radial guiding-center drifts [1]. A magnetic field is termed omnigenous [1,2] if the radial drift of all particles averages to zero in time, where the average is performed over the leading-order motion along a field line. As particles conserve the longitudinal adiabatic invariant $J = \oint v_\parallel \, d\ell$ during their drift motion, omnigeneity can be defined equivalently as the constancy of $J$ over each flux surface. Omnigeneity is guaranteed for a field which is quasisymmetric [3,4], meaning that $B$ varies on a flux surface only through a fixed linear combination of the Boozer angles [5]. The conserved quantity associated with this symmetry implies that all orbits are strictly confined. A field can never be exactly quasisymmetric [6], but fields can be found which are approximately so [4,7]. However, requiring even approximate quasisymmetry places a severe constraint on stellarator optimization. It is therefore desirable to examine non-quasisymmetric fields which are still nearly omnigenous. In [8,9] it was shown that any analytic and perfectly omnigenous field must be quasisymmetric, but analytic fields can be constructed which are nearly omnigenous and yet are far from being quasisymmetric.

Omnigenous fields have many remarkable mathematical and physical properties. For example, it is proven in [8] that the maximum and minimum of $B$ are the same for each field line on a flux surface. In contrast to a general stellarator, the vanishing of the time-averaged radial drift implies that no $D \propto 1/\nu$ regime exists in a perfectly omnigenous device, where $D$ is any radial transport coefficient and $\nu$ is the relevant collision frequency.

An important subset of omnigeneous fields is the set of fields which are quasi-isodynamic [10-13]. A field is defined as being quasi-isodynamic [13] if it is omnigenous and if the constant-$B$ contours close poloidally, as opposed to closing toroidally or helically. Quasi-isodynamic fields are experimentally relevant, as the W7-X stellarator is designed to be approximately quasi-isodynamic at high beta [14,15], and recent stellarator design studies have examined equilibria which are even closer to perfect quasi-isodynamicity [11]. Axisymmetric fields are not quasi-isodynamic since constant-$B$ contours close toroidally. In a quasi-isodynamic plasma, the part of the long-mean-free-path (banana) regime distribution function determined by the collisional constraint is found from an equation which is identical in form to the analogous equation for an axisymmetric plasma [13]. Also, if there is no net toroidal current inside a flux surface, then the bootstrap current vanishes [11,13].

Several new results for quasi-isodynamic fields are obtained in the following sections. In section 2, we give new derivations for some of the properties of quasi-isodynamic fields discussed in [13], and new expressions are also derived which relate the components of **B** to derivatives of $B$ and of the Boozer angles. In section 3, we use these relations to derive novel expressions for the flow and current in a quasi-isodynamic field. It is shown that the distribution function obtained in [13] for the long-mean-free-path regime is consistent with these forms, but our forms of the flow and current are valid for all regimes of collisionality. In section 4, we use ambipolarity to determine the radial electric field in the long-mean-free-path regime. The bootstrap current is given in section 5, and we conclude in section 6. Emphasis is given throughout on practical calculations for a specified $B(\theta, \zeta)$, and the figures show an example calculation of the key quantities for a model field.

Flows and transport in a general stellarator have been discussed previously in many references, such as [16-20]. Here we focus on quasi-isodynamic devices to highlight the additional analytic results which can be obtained due to the strong constraint on the field geometry. Although the field of any real stellarator will not be perfectly quasi-isodynamic, the results which follow are useful in several regards. A perfectly quasi-isodynamic model field can be provided as input to transport codes, so the results herein can be used to validate such codes. In addition, the flow, current, and radial electric field in a plasma which is approximately quasi-isodynamic can be expected to resemble the analytic forms derived here. Therefore, the results herein may give insight into the physics of W7-X-like stellarators.

## 2. Geometric properties of quasi-isodynamic fields

We begin by writing the magnetic field in Boozer coordinates [5]:

$$\mathbf{B} = \nabla \psi \times \nabla \theta + q^{-1} \nabla \zeta \times \nabla \psi \tag{1}$$

and

$$\mathbf{B} = I \nabla \zeta + K \nabla \theta + L \nabla \psi, \tag{2}$$

where $2\pi\psi$ is the toroidal flux, $cI(\psi)/2$ is the poloidal current linked outside the flux surface, $cK(\psi)/2$ is the toroidal current inside the flux surface, and $q(\psi)$ is the safety factor. We introduce the field line label $\alpha = \theta - q^{-1}\zeta$, so $\mathbf{B} = \nabla\psi \times \nabla\alpha$. Let $\breve{B}$ and $\hat{B}$ denote the minimum and maximum of $B$ along a field line, respectively. Omnigeneity requires that the most deeply trapped particles (those with $v_\parallel = 0$) have no radial drift, so if the potential is a flux function, the radial $\nabla B$ drift $\propto \mathbf{B} \times \nabla B \cdot \nabla \psi$ must vanish. As these particles also lie at $B = \breve{B}$ where $\mathbf{B} \cdot \nabla B = 0$, then $\breve{B}$ must be independent of field line ($\partial \breve{B} / \partial \alpha = 0$). By considering the action of marginally trapped particles, it is proven in [8] that $\hat{B}$ must be independent of field line as well. As the range of allowed $B$ is thus independent of $\alpha$, and as a $2\pi$ increase in $\alpha$ at fixed $B$ is a closed poloidal loop, it becomes convenient to use $(\psi, \alpha, B)$ as coordinates. However, specifying $(\psi, \alpha, B)$ does not uniquely determine a location on the flux surface – there are two possible locations in each magnetic well, one on either side of $\breve{B}$ along the field line. We denote this discrete degree of freedom by $\gamma = \pm 1$, which Helander and Nührenberg term the *branch* [13].

In an omnigenous field, the longitudinal adiabatic invariant $J = \oint v_\| \, d\ell$ must be constant on a flux surface. This requirement implies [8]

$$\left(\frac{\partial}{\partial \alpha}\right)_B \sum_\gamma \frac{\gamma}{\mathbf{b} \cdot \nabla B} = 0, \tag{3}$$

a result which is termed the "Cary-Shasharina Theorem" in [13]. Here and throughout, subscripts on partial derivatives indicate which quantities are held fixed. A proof of (3) is given in appendix A. It is shown in [8] that (3) implies the contours of $B = \hat{B}$ are straight in Boozer coordinates. In a quasi-isodynamic field, the $B$ contours close poloidally, and so $B = \hat{B}$ contours must be curves of constant $\zeta$.

It is useful to express $\mathbf{B}$ in terms of its covariant components in the $(\psi, \alpha, B)$ coordinates:

$$\mathbf{B} = B_\psi \nabla \psi + B_\alpha \nabla \alpha + B_B \nabla B. \tag{4}$$

From the dot product of this expression with $\mathbf{B} = \nabla \psi \times \nabla \alpha$ we find

$$B_B = \frac{B^2}{\mathbf{B} \cdot \nabla B}. \tag{5}$$

The dot product of (4) with $\nabla \psi \times \nabla B$ gives $B_\alpha = -\mathbf{B} \times \nabla \psi \cdot \nabla B / \mathbf{B} \cdot \nabla B$, and the dot product of (4) with $\nabla \alpha \times \nabla B$ gives $B_\psi = \mathbf{B} \times \nabla \alpha \cdot \nabla B / \mathbf{B} \cdot \nabla B$.

We note that an incremental step can be written in the $(\psi, \alpha, B)$ coordinates as

$$d\mathbf{r} = \frac{1}{\mathbf{B} \cdot \nabla B}\left[\left(\nabla \alpha \times \nabla B\right)d\psi + \left(\nabla B \times \nabla \psi\right)d\alpha + \mathbf{B}\,dB\right] \tag{6}$$

and in the $(\psi, \theta, \zeta)$ coordinates by

$$d\mathbf{r} = \frac{1}{\mathbf{B} \cdot \nabla \zeta}\left[\left(\nabla \theta \times \nabla \zeta\right)d\psi + \left(\nabla \zeta \times \nabla \psi\right)d\theta + \left(\nabla \psi \times \nabla \theta\right)d\zeta\right]. \tag{7}$$

By applying Ampere's law to a constant-$B$ poloidal loop on a flux surface, and using (6) to write $\mathbf{B} \cdot d\mathbf{r} = B_\alpha d\alpha$ for this path, we find $\int_0^{2\pi} B_\alpha \, d\alpha = 2\pi K$, where $K(\psi)$ is the same quantity as in (2). It follows that [13]

$$B_\alpha = -\frac{\mathbf{B} \times \nabla \psi \cdot \nabla B}{\mathbf{B} \cdot \nabla B} = K + \left(\frac{\partial h}{\partial \alpha}\right)_B \tag{8}$$

for some single-valued function $h$. As shown in [21], $\mathbf{B} \times \nabla \psi \cdot \nabla B / \mathbf{B} \cdot \nabla B$ is a flux function if and only if the field is quasisymmetric, so the $(\partial h / \partial \alpha)_B \to 0$ limit corresponds to quasisymmetry. More precisely, $(\partial h / \partial \alpha)_B \to 0$ corresponds to quasi-poloidal symmetry ($B = B(\psi, \zeta)$) since, by definition, $B$ contours in a quasi-isodynamic field close poloidally.

Next, we use the fact that $(\nabla \times \mathbf{B}) \times \mathbf{B}$ is parallel to $\nabla \psi$ in a scalar-pressure MHD equilibrium, so

$$0 = (\nabla \times \mathbf{B}) \cdot \nabla \psi = \nabla \cdot (\mathbf{B} \times \nabla \psi) = (\mathbf{B} \cdot \nabla B) \left[ \left( \frac{\partial B_B}{\partial \alpha} \right)_B - \left( \frac{\partial B_\alpha}{\partial B} \right)_\alpha \right]. \tag{9}$$

Plugging in (5) and (8), we obtain the useful identity

$$\left( \frac{\partial}{\partial \alpha} \right)_B \frac{B^2}{\mathbf{B} \cdot \nabla B} = \frac{\partial^2 h}{\partial \alpha \partial B}. \tag{10}$$

Applying (3), $\sum_\gamma \gamma\, \partial^2 h / \partial \alpha \partial B = 0$. Integrating this result from $\breve{B}$ to $B$, then $\sum_\gamma \gamma (\partial h / \partial \alpha)_B = 0$, i.e. $(\partial h / \partial \alpha)_B$ must be branch-independent everywhere. In this integration, the contribution from the $\breve{B}$ boundary vanishes because $\gamma = +1$ and $\gamma = -1$ refer to the same location there, so $B_\alpha$ is branch-independent there, and so $(\partial h / \partial \alpha)_B$ is branch-independent there as well. Although $(\partial h / \partial \alpha)_B$ must therefore be $\gamma$-independent everywhere, $h$ itself may depend on the branch. However, consider the quantity $h_\Sigma = [h(\gamma=1) + h(\gamma=-1)]/2$. By applying $(\partial / \partial \alpha)_B$ to $h(\gamma) = h_\Sigma + \tfrac{1}{2}[h(\gamma) - h(-\gamma)]$, we then find $(\partial h / \partial \alpha)_B = (\partial h_\Sigma / \partial \alpha)_B$, so $h$ could be replaced by $h_\Sigma$ in (8), the equation which defined $h$. Thus, it is no loss in generality to assume $h$ is branch-independent. This proof differs somewhat from the one given in [13], though the underlying principles are the same. In the sections which follow, we will only ever need $(\partial h / \partial \alpha)_B$ rather than $h$ itself, so the calculation of $h_\Sigma$ will not be necessary.

For completeness, we now present one additional relation which can be found for $B_\psi$. We use (4) to form

$$(\nabla \times \mathbf{B}) \times \mathbf{B} = (\mathbf{B} \cdot \nabla B) \left[ \left( \frac{\partial B_\psi}{\partial B} \right)_\alpha - \left( \frac{\partial}{\partial \psi} \right)_{\alpha, B} \left( \frac{B^2}{\mathbf{B} \cdot \nabla B} \right) \right] \nabla \psi. \tag{11}$$

Then, from MHD equilibrium $4\pi \nabla p = (\nabla \times \mathbf{B}) \times \mathbf{B}$ it follows that

$$\left( \frac{\partial B_\psi}{\partial B} \right)_\alpha = \left( \frac{\partial}{\partial \psi} \right)_{\alpha, B} \left( \frac{B^2}{\mathbf{B} \cdot \nabla B} \right) + \frac{4\pi}{\mathbf{B} \cdot \nabla B} \frac{dp}{d\psi}. \tag{12}$$

We now relate the components of $\mathbf{B}$ in the $(\psi, \alpha, B)$ basis to the more familiar quantities $\theta$ and $\zeta$. The results will be important in later sections for calculating the parallel flow and current. First, the dot product of (1) with (2) gives $\mathbf{B} \cdot \nabla \theta = B^2 / (qI + K) = q^{-1} \mathbf{B} \cdot \nabla \zeta$, and it follows that

$$\frac{B^2}{\mathbf{B} \cdot \nabla B} = (qI + K) \left[ \left( \frac{\partial B}{\partial \theta} \right)_\zeta + q \left( \frac{\partial B}{\partial \zeta} \right)_\theta \right]^{-1}. \tag{13}$$

Using $(\partial B / \partial \theta)_\alpha = (\partial B / \partial \theta)_\zeta + q (\partial B / \partial \zeta)_\theta$, then

$$\frac{B^2}{\mathbf{B} \cdot \nabla B} = (qI + K) \left( \frac{\partial \theta}{\partial B} \right)_\alpha = \frac{qI + K}{q} \left( \frac{\partial \zeta}{\partial B} \right)_\alpha. \tag{14}$$

Also, from (2) we can form

$$\mathbf{B} \times \nabla \psi \cdot \nabla B = (\mathbf{B} \cdot \nabla \theta) q \left[ I \left( \frac{\partial B}{\partial \theta} \right)_\zeta - K \left( \frac{\partial B}{\partial \zeta} \right)_\theta \right]. \tag{15}$$

Plugging this result into (8) we obtain

$$\left(\frac{\partial h}{\partial \alpha}\right)_B = -(qI+K)\left(\frac{\partial B}{\partial \theta}\right)_\zeta \left[\left(\frac{\partial B}{\partial \theta}\right)_\zeta + q\left(\frac{\partial B}{\partial \zeta}\right)_\theta\right]^{-1}. \tag{16}$$

Using $(\partial B/\partial \theta)_\zeta = (\partial B/\partial \theta)_\alpha + (\partial B/\partial \alpha)_\theta$, then (16) implies

$$\left(\frac{\partial h}{\partial \alpha}\right)_B = (qI+K)\left[\left(\frac{\partial \theta}{\partial \alpha}\right)_B - 1\right] = \frac{qI+K}{q}\left(\frac{\partial \zeta}{\partial \alpha}\right)_B \tag{17}$$

where we have used

$$0 = \left(\frac{\partial B}{\partial \theta}\right)_B = \left(\frac{\partial B}{\partial \theta}\right)_\alpha + \left(\frac{\partial \alpha}{\partial \theta}\right)_B \left(\frac{\partial B}{\partial \alpha}\right)_\theta. \tag{18}$$

Comparing (14) and (17), it is evident that (10) is satisfied. Equations (16) and (17) are key results we will need for computations in section 3.

As the $B = \hat{B}$ contours are constant-$\zeta$ curves, then $(\partial \zeta/\partial \alpha)_B = 0$ along these contours. Equation (17) then implies $(\partial h/\partial \alpha)_B \to 0$ along these contours as well.

Let $\Delta \zeta(B) = \sum_\gamma \gamma \int_{\breve{B}}^B (\partial \zeta/\partial B)_\alpha d\dot{B}$ be the difference in $\zeta$ between the two points of field magnitude $B$ on either side of $\breve{B}$, where $\zeta = \zeta(B)$. By (14) and (3),

$$(\partial(\Delta \zeta)/\partial \alpha)_B = 0 \tag{19}$$

for a quasi-isodynamic field. In section IV of [8], Cary and Shasharina give a procedure to construct a function $B(\theta, \zeta)$ with the property (19) and with $\hat{B}$ straight in Boozer coordinates. This construction is reviewed briefly in appendix B since we will use it to obtain the figures. Differentiating (19), then

$$\sum_\gamma \gamma \frac{\partial^2 \zeta}{\partial \alpha \partial B} = 0 \tag{20}$$

which, due to (14), implies (3). Thus, any field generated by the Cary-Shasharina construction will be quasi-isodynamic, in the sense that all the properties described in this section will apply. For example, $(\partial h/\partial \alpha)_B$ can be calculated from (16), and the result will automatically be branch-independent. (This last fact can also be seen by integrating (20) from $\breve{B}$ and using (17).)

Figure 1 shows a quasi-isodynamic field generated using the Cary-Shasharina construction, with parameters specified in appendix B. The property (19) is illustrated in figure 1 by the fact that the two thick line segments, both of which are parallel to the field lines, have the same $\Delta \zeta$. Figure 2 shows $(\partial h/\partial \alpha)_B$ which is calculated for this field using (16).

## 3. Parallel flows and current

*3.a. General collisionality case*

We now derive the form of the flow for any particle species in a quasi-isodynamic field. We assume the species density $n$, pressure $p$, and electrostatic potential $\Phi$ are all flux functions to leading order. The perpendicular flow is given to leading order by the sum of the $\mathbf{E}\times\mathbf{B}$ and diamagnetic flows: $\mathbf{V}_\perp = \omega B^{-2}\mathbf{B}\times\nabla\psi$, where $\omega(\psi) = c\Phi' + cp'/(Zen)$, $Ze$ is the species charge, and primes denote $\partial/\partial\psi$. Writing the total flow as $\mathbf{V} = \mathbf{V}_\perp + V_\parallel B^{-1}\mathbf{B}$, then from the mass conservation relation $\nabla\cdot(n\mathbf{V}) = 0$ we can write

$$\mathbf{B}\cdot\nabla\left(\frac{V_\parallel + \omega U}{B}\right) = 0 \tag{21}$$

where $U$ is defined to be a single-valued and continuous solution of

$$\mathbf{B}\cdot\nabla(U/B) = \mathbf{B}\times\nabla\psi\cdot\nabla(1/B^2). \tag{22}$$

The solvability condition of (22) is satisfied for any scalar-pressure MHD equilibrium. Applying (8), we find

$$\left(\frac{\partial(U/B)}{\partial B}\right)_\alpha = \frac{2}{B^3}\left[K + \left(\frac{\partial h}{\partial\alpha}\right)_B\right]. \tag{23}$$

Integrating in $B$,

$$Y = \frac{U}{B} + \frac{K+W}{B^2} \tag{24}$$

where $Y$ is the integration constant, and

$$W(\psi,\,\alpha,\,B) = 2B^2\int_B^{\hat B}\frac{d\dot B}{\dot B^3}\left(\frac{\partial h}{\partial\alpha}\right)_{\dot B}. \tag{25}$$

Again, dots underneath a quantity indicate it is evaluated using the dummy integration variable $\dot B$ rather than using the local value of $B$. We will see the quantity $W$ plays an important role in quasi-isodynamic fields.

While $Y$ is by definition independent of $B$, additional work is required to show that $Y$ is also independent of $\alpha$ and $\gamma$, which we do as follows. At $B = \breve B$, both signs of $\gamma$ refer to the same location, so $U$ is $\gamma$-independent there. Consequently the entire right-hand side of (24) is $\gamma$-independent at $B = \breve B$, so $Y$ must be $\gamma$-independent. Next, the right-hand side of (24) is continuous across the curve $B = \hat B$ when $\theta$ is held fixed. Therefore $Y$ must have this same property, implying $Y(\alpha) = Y(\alpha + 2\pi/(qN))$ for all $\alpha$, where again $N$ is the number of identical stellarator cells. It follows that $Y$ must be independent of $\alpha$, and therefore $Y$ is a flux function. Had $\breve B$ been chosen as the limit of integration in (25) instead of $\hat B$, then the right-hand side of (24) would not be continuous across the curve $B = \hat B$ when $\theta$ is held fixed, and so $Y$ would need to depend on $\alpha$.

We choose to specify $Y$ by requiring $\langle UB\rangle = 0$, where the brackets denote a flux surface average. For any quantity $Q$, this average is

$$\langle Q \rangle = \frac{\sum_\gamma \gamma \int_{\check{B}}^{\hat{B}} dB \int_0^{2\pi} d\alpha \frac{Q}{\mathbf{B} \cdot \nabla B}}{\sum_\gamma \gamma \int_{\check{B}}^{\hat{B}} dB \int_0^{2\pi} d\alpha \frac{1}{\mathbf{B} \cdot \nabla B}}. \tag{26}$$

Note that $\langle W \rangle = 0$, since the $\alpha$ integral can be performed by parts, and (3) applied. We thereby obtain

$$U = -\frac{1}{B}\left[\left(1 - \frac{B^2}{\langle B^2 \rangle}\right)K + W\right]. \tag{27}$$

Next, it follows from (21) that $V_\| B + \omega UB = B^2 A$ for some flux function $A(\psi)$. Flux-surface-averaging this equation to find $A$, we obtain our final form for the parallel flow of each species in a quasi-isodynamic field:

$$V_\| = \frac{B \langle V_\| B \rangle}{\langle B^2 \rangle} + V_\|^{\text{PS}}, \tag{28}$$

where the "Pfirsch-Schlüter flow" is

$$V_\|^{\text{PS}} = \frac{c}{B}\left(\Phi' + \frac{p'}{Zen}\right)\left[\left(1 - \frac{B^2}{\langle B^2 \rangle}\right)K + W\right] \tag{29}$$

and satisfies $\langle V_\|^{\text{PS}} B \rangle = 0$. For $(\partial h/\partial \alpha)_B \to 0$, then $W \to 0$, and (29) reduces properly to the result for a quasi-poloidally symmetric field.

The parallel current has a similar form to (28)-(29), which can be obtained by multiplying these equations by $Zen$ and summing over species. The result is

$$j_\| = \frac{B \langle j_\| B \rangle}{\langle B^2 \rangle} + j_\|^{\text{PS}}, \tag{30}$$

where the "Pfirsch-Schlüter current" is

$$j_\|^{\text{PS}} = \frac{cp'_{\text{tot}}}{B}\left[\left(1 - \frac{B^2}{\langle B^2 \rangle}\right)K + W\right] \tag{31}$$

and satisfies $\langle j_\|^{\text{PS}} B \rangle = 0$, and $p_{\text{tot}}$ is the sum of the pressures of each species.

We emphasize that (28)-(31) are valid in a quasi-isodynamic stellarator at arbitrary collisionality.

*3.b. Long-mean-free-path regime*

We now show that the distribution function obtained in [13] for the long-mean-free-path regime indeed gives a flow of the form (28)-(29). The distribution function obtained in that reference was

$$f = f_0 - f_0 \frac{e\Phi_1}{T} + g + K \frac{v_\parallel}{\Omega} \frac{\partial f_0}{\partial \psi} - \frac{cm}{Ze} \int_B^{B_x} d\dot{B} \left( \frac{\partial \dot{h}}{\partial \alpha} \right)_B \frac{\partial}{\partial \dot{B}} \frac{v_\parallel}{\dot{B}} \frac{\partial f_0}{\partial \psi} \tag{32}$$

where $f_0$ is a stationary Maxwellian flux function, $\Phi_1(\psi, \theta, \zeta)$ is the next-order correction to the potential, and $\Omega = ZeB/(mc)$ is the gyrofrequency. The $\partial/\partial \psi$ derivatives and the $d\dot{B}$ integral are performed holding the leading-order energy $v^2/2 + Ze\Phi/m$ and the magnetic moment $v_\perp^2/2B$ fixed. Also,

$$B_x = \begin{cases} \hat{B} & \text{if } \lambda < 1/\hat{B} \\ 1/\lambda & \text{if } \lambda > 1/\hat{B}, \end{cases} \tag{33}$$

where $\lambda = v_\perp^2/(Bv^2)$, and $g$ is a flux function which vanishes for trapped particles. Applying $\int d^3v\, v_\parallel$ to (32) gives

$$nV_\parallel = X + \frac{cnK}{B}\left(\Phi' + \frac{p'}{Zen}\right)$$
$$- \frac{cm}{Ze} \int d^3v\, v_\parallel f_0 \left[ \frac{p'}{p} + \frac{Ze\Phi'}{T} + \left(\frac{mv^2}{2T} - \frac{5}{2}\right)\frac{T'}{T} \right] \int_B^{B_x} d\dot{B} \left(\frac{\partial \dot{h}}{\partial \alpha}\right)_B \frac{\partial}{\partial \dot{B}} \frac{v_\parallel}{\dot{B}} \tag{34}$$

where $X = \int d^3v\, v_\parallel g$. We can write $X = \pi B \sum_\sigma \sigma \int_0^\infty dv\, v^3 \int_0^{1/\hat{B}} d\lambda\, g$, where $\sigma = \text{sgn}(v_\parallel)$, and the upper limit of the $\lambda$ integral is changed from $1/B$ to $1/\hat{B}$ since $g$ is zero for trapped particles. As $g$ is a flux function at fixed $v$ and $\lambda$, then $X \propto B$, so $X$ has the form of the $B\langle V_\parallel B\rangle/\langle B^2\rangle$ term in (28). Finally, we evaluate the last line of (34), obtaining as a first step

$$\int d^3v\, v_\parallel f_0 \int_B^{B_x} d\dot{B} \left(\frac{\partial \dot{h}}{\partial \alpha}\right)_B \frac{\partial}{\partial \dot{B}} \frac{v_\parallel}{\dot{B}} = -\frac{3nTB}{4m} \int_0^{1/B} d\lambda \int_B^{B_x} d\dot{B} \left(\frac{\partial \dot{h}}{\partial \alpha}\right)_B \frac{(2-\lambda\dot{B})}{\dot{B}^2 \sqrt{1-\lambda\dot{B}}}. \tag{35}$$

The contribution from the $T'$ term in (34) vanishes in the $v$ integration. We now switch the order of the $\lambda$ and $\dot{B}$ integrations, so $\dot{B}$ ranges over $(B, \hat{B})$ and $\lambda$ ranges over $(0, 1/\dot{B})$. The $\lambda$ integral can then be evaluated, giving precisely the $W$ term in (29). Thus, the flow associated with the distribution (32) indeed has the form (28)-(29).

As described in [13], the $g$ component of the distribution function for passing particles is computed from the constraint $\langle (B/v_\parallel)C \rangle = 0$, where $C$ is the collision operator. If we consider the ions in a pure plasma, we can explicitly calculate $g_i$ and $X$ using the standard momentum-conserving model operator

$$C_i = \nu \mathcal{L}\left\{ f_{i1} - f_{i0}\frac{m_i u v_\parallel}{T_i}\right\} \tag{36}$$

where

$$\mathcal{L} = \frac{2v_\parallel}{v^2 B} \frac{\partial}{\partial \lambda} \lambda v_\parallel \frac{\partial}{\partial \lambda} \tag{37}$$

is the Lorentz pitch-angle scattering operator,

$$u = \left[ \int d^3v\, f_{i0} \frac{m_i v^2}{3T_i} \nu \right]^{-1} \int d^3v\, f_{i1} \nu v_\parallel, \tag{38}$$

$$\nu = \frac{2\pi Z^4 e^4 n_i \ln \Lambda}{\sqrt{2m_i} T_i^{3/2}} \frac{\left[\mathrm{erf}(x) - \Psi(x)\right]}{x^3}, \tag{39}$$

$\Psi(x) = \left[\mathrm{erf}(x) - x\,\mathrm{erf}'(x)\right]/(2x^2)$, $\mathrm{erf}(x) = (2/\sqrt{\pi}) \int_0^x \exp(-t^2) dt$ is the error function, and $x = \upsilon / \sqrt{2T_i/m_i}$. The analysis then closely resembles the standard tokamak calculation [1], giving

$$g_i = -f_{i0} \frac{Km_i c}{Ze} \frac{T_i'}{T_i} \left[\frac{m_i \upsilon^2}{2T_i} - 1.33\right] \frac{\upsilon}{2} H \int_\lambda^{\hat{B}^{-1}} \frac{d\lambda}{\left\langle\sqrt{1-\lambda B}\right\rangle} \tag{40}$$

where $H = H\left(\hat{B}^{-1} - \lambda\right)$ is a Heavyside function which is 1 for passing particles and 0 for trapped particles. The final expression for the parallel ion flow in the long-mean-free-path regime becomes

$$V_{i\|} = -1.17 f_c \frac{KcBT_i'}{Ze\langle B^2\rangle} + \frac{c}{B}\left(\Phi' + \frac{p_i'}{Zen_i}\right)(K+W) \tag{41}$$

where the effective fraction of circulating particles

$$f_c = \frac{3}{4}\langle B^2\rangle \int_0^{\hat{B}^{-1}} \frac{\lambda\, d\lambda}{\left\langle\sqrt{1-\lambda B}\right\rangle} \tag{42}$$

is approximately one if the variation in $B$ is small. As expected, (41) has the form (28)-(29).

Notice that the distribution function (32) was derived in [13] only for the long-mean-free-path regime, whereas the general form of the flow (28)-(29) was derived here without any assumption about the collisionality.

*3.c. Computation of geometric integral*

For any particular quasi-isodynamic field, the $W$ quantity (25) which appears in the Pfirsch-Schlüter flow (29) and Pfirsch-Schlüter current (31) can be evaluated numerically in a number of ways. One approach is to use (17) to write

$$W = 2B^2 \frac{qI+K}{q} \int_B^{\hat{B}} \frac{dB}{B^3} \left(\frac{\partial \zeta}{\partial \alpha}\right)_B. \tag{43}$$

The function $\zeta(\alpha, B)$ can be computed from a given field $B(\theta, \zeta)$ or specified directly. The integrand in (43) can then be computed, the integral evaluated, and if desired, the result mapped to $(\theta, \zeta)$ coordinates. Figure 3 shows $W$ computed using this method for the model field of figure 1.

In an alternative approach, we begin by transforming (25) as follows:

$$W = -2B^2 \int_{\zeta_x}^{\zeta} \frac{d\zeta}{B^3}\left(\frac{\partial B}{\partial \zeta}\right)_\alpha \left(\frac{\partial h}{\partial \alpha}\right)_B = -\frac{2B^2}{q}\int_{\zeta_x}^{\zeta} \frac{d\zeta}{B^3}\left[\left(\frac{\partial B}{\partial \theta}\right)_\zeta + q\left(\frac{\partial B}{\partial \zeta}\right)_\theta\right]\left(\frac{\partial h}{\partial \alpha}\right)_B. \tag{44}$$

For points between $\breve{B}$ and $\zeta = 0$, the integration bound $\zeta_x$ is 0, and for points on the other side of $\breve{B}$, $\zeta_x = 2\pi/N$, where again $N$ is the number of toroidal periods of the stellarator (i.e. the number of $\hat{B}$ or $\breve{B}$ curves). Applying (16), then

$$W = 2B^2 \frac{qI+K}{q} \int_{\zeta_x}^{\zeta} \frac{d\zeta}{\dot{B}^3} \left(\frac{\partial B}{\partial \theta}\right)_\zeta. \tag{45}$$

This expression can also be obtained from (43) using (18), with $\theta$ replaced by $\zeta$ in the latter. Depending on the particular application, either the form (43) or (45) for $W$ may be more convenient to evaluate.

Notice that the integrals throughout this section are performed along constant-$\alpha$ paths.

As the right-hand side of (25) is branch-independent, then the integrals in (43) and (45) must be so as well. These integrals can still be computed for a $B(\theta, \zeta)$ which is not quasi-isodynamic, but the result will be discontinuous at $B = \breve{B}$. For a *nearly* quasi-isodynamic $B(\theta, \zeta)$, such as the one obtained in [8,9] by Fourier-filtering a perfectly quasi-isodynamic field, the discontinuity in $W$ is small.

## 4. Particle flux and radial electric field

We now derive an expression for the radial flux of each particle species at any collisionality. The neoclassical part of the flux is

$$\langle \mathbf{\Gamma} \cdot \nabla \psi \rangle = \left\langle \int d^3 v \, f \mathbf{v}_d \cdot \nabla \psi \right\rangle \tag{46}$$

where $\mathbf{v}_d = (v_\parallel / \Omega) \nabla \times (v_\parallel \mathbf{b})$ is the drift velocity. As in section 3b, derivatives throughout this section will hold the leading-order energy $v^2/2 + Ze\Phi/m$ and the magnetic moment $v_\perp^2/2B$ fixed. It can be shown using (8) that $\mathbf{v}_d \cdot \nabla \psi = v_\parallel \mathbf{b} \cdot \nabla \Delta$ where

$$\Delta = -K \frac{v_\parallel}{\Omega} + S \tag{47}$$

and

$$S = \int_B^{B_x} d\dot{B} \left(\frac{\partial h}{\partial \alpha}\right)_B \frac{\partial}{\partial \dot{B}} \frac{v_\parallel}{\Omega}. \tag{48}$$

Using

$$\int d^3 v (\,\cdot\,) = \pi B \sum_\sigma \sigma \int_0^\infty v^2 dv \int_0^{1/B} \frac{d\lambda}{\sqrt{1-\lambda B}} (\,\cdot\,) \tag{49}$$

and the fact that $\langle \mathbf{B} \cdot \nabla Q \rangle = 0$ for any single-valued $Q$, we find

$$\langle \mathbf{\Gamma} \cdot \nabla \psi \rangle = \left\langle \int d^3 v \, f v_\parallel \mathbf{b} \cdot \nabla \Delta \right\rangle = -\left\langle \int d^3 v \, v_\parallel \Delta \mathbf{b} \cdot \nabla f \right\rangle. \tag{50}$$

In the last equation we have also used the fact that $\Delta = 0$ at $\lambda = 1/B$. Substituting in the drift-kinetic equation

$$v_\parallel \mathbf{b} \cdot \nabla f + \mathbf{v}_d \cdot \nabla \psi \frac{\partial f_0}{\partial \psi} = C \tag{51}$$

where $C$ is the collision operator, we then obtain

$$\langle \Gamma \cdot \nabla \psi \rangle = -\left\langle \int d^3 \upsilon \, \Delta C \right\rangle. \tag{52}$$

We now specialize to consider the flux of ions in a pure plasma, so ion collisions with electrons and the radial electron particle flux can be ignored to leading order in $m_e / m_i$. We also specialize to the long-mean-free-path collisionality regime. The model collision operator (36) is again employed. This operator is constructed to have the momentum conservation property $\int d^3 \upsilon \, \upsilon_\| C = 0$, so (52) can be written

$$\langle \Gamma_i \cdot \nabla \psi \rangle = -\left\langle \int d^3 \upsilon \, S \nu \mathcal{L} \left\{ -\Delta \frac{\partial f_{i0}}{\partial \psi} + g_i - f_{i0} \frac{m_i u \upsilon_\|}{T_i} \right\} \right\rangle \tag{53}$$

where we have applied the distribution function (32).

We next make use of the property $\langle (\partial Q / \partial \alpha)_B \rangle = 0$ for any branch-independent $Q$. This property follows from (26) and (10) if an integration by parts is performed in $\alpha$. It follows that any terms in (53) which are linear in $(\partial h / \partial \alpha)_B$ will vanish. For example, $g_i$ is a flux function and so it does not contribute to (53). We group the nonvanishing terms into two pieces as follows:

$$\langle \Gamma_i \cdot \nabla \psi \rangle = \Gamma_1 + \Gamma_2 \tag{54}$$

where

$$\Gamma_1 = \left\langle \int d^3 \upsilon \frac{\partial f_{i0}}{\partial \psi} S \nu \mathcal{L} \{S\} \right\rangle, \tag{55}$$

$$\Gamma_2 = \left\langle \int d^3 \upsilon \, S \nu \mathcal{L} \left\{ f_{i0} \frac{m_i u_\alpha \upsilon_\|}{T_i} \right\} \right\rangle, \tag{56}$$

and

$$u_\alpha = -\left[ \int d^3 \upsilon \, f_{i0} \frac{m_i \upsilon^2}{3 T_i} \nu \right]^{-1} \int d^3 \upsilon \, \frac{\partial f_{i0}}{\partial \psi} \nu \upsilon_\| S \tag{57}$$

is the part of (38) which depends on $\alpha$. Notice that upon applying (49), the $d^3 \upsilon$ velocity integral in $\Gamma_1$ gives

$$\Gamma_1 \propto \int_0^\infty d\upsilon \frac{\partial f_{i0}}{\partial \psi} \upsilon^4 \nu. \tag{58}$$

Also, examining (57), $u_\alpha$ is proportional to the same factor. Therefore $\Gamma_2$ and the total flux have the same proportionality:

$$\langle \Gamma_i \cdot \nabla \psi \rangle \propto \int_0^\infty d\upsilon \frac{\partial f_{i0}}{\partial \psi} \upsilon^4 \nu. \tag{59}$$

Consequently, when $(\partial h / \partial \alpha)_B$ is nonzero, the $\upsilon$ integrals can be factored out of the ambipolarity condition $\langle \Gamma_i \cdot \nabla \psi \rangle \approx 0$ to obtain

$$0 = \int_0^\infty dx \, x^4 e^{-x^2} \nu \left[ \frac{p_i'}{p_i} + \frac{Ze\Phi'}{T_i} + \left( x^2 - \frac{5}{2} \right) \frac{T_i'}{T_i} \right]. \tag{60}$$

We can rearrange to solve for the radial electric field, using

$$\frac{5}{2} - \left[\int_0^\infty dx\, x^4 e^{-x^2} \nu\right]^{-1} \int_0^\infty dx\, x^6 e^{-x^2} \nu = 1.17, \tag{61}$$

a number which is familiar from the banana-regime analysis in a tokamak and from (41). The radial electric field in a long-mean-free-path-regime quasi-isodynamic stellarator is therefore

$$\frac{Ze\Phi'}{T_i} = -\frac{p_i'}{p_i} + 1.17 \frac{T_i'}{T_i}. \tag{62}$$

Physically, this formula shows an inward electric field $\mathbf{E} \approx T_i \nabla n_i / (Zen_i)$ is required to reduce the ion flux down to the level of the electron flux, corresponding to "ion root" confinement.

Note that if a pitch-angle scattering operator without the momentum conserving term $u$ was used in the above calculation, the same radial electric field would be obtained.

The result (62) can be used to simplify (41), leaving the parallel ion flow in a long-mean-free-path-regime quasi-isodynamic stellarator as

$$V_{\|i} = 1.17 \frac{cT_i'}{ZeB} \left[\left(1 - f_c \frac{B^2}{\langle B^2 \rangle}\right) K + W\right]. \tag{63}$$

Notice that $V_{i\|} \propto T_i'$ so an ion temperature gradient is required for there to be a parallel ion flow.

**5. Bootstrap current**

In a quasi-isodynamic field, the bootstrap current can be calculated exactly as in a tokamak [13], for although the $(\partial h/\partial\alpha)_B$ term in the distribution function (32) does not arise in a tokamak, this term vanishes whenever a flux surface average is taken, as argued following (53). Therefore in the evaluation of $\langle j_\| B \rangle$ using the standard analytical method for a tokamak [1], the $(\partial h/\partial\alpha)_B$ terms disappear from the analysis.

For the long-mean-free-path regime and arbitrary ion charge $Z$ the result is

$$\langle j_\| B \rangle = f_t Kc \frac{Z^2 + 2.21Z + 0.75}{Z(\sqrt{2} + Z)} \left(p_i' + p_e' - \frac{2.07Z + 0.88}{Z^2 + 2.21Z + 0.75} n_e T_e' - 1.17 n_e \frac{T_i'}{Z}\right), \tag{64}$$

where $f_t = 1 - f_c$ is the effective trapped fraction and $f_c$ is given by (42). To obtain (64) we have used the Spitzer function as described on page 207 of [1], and we have used the approximate Spitzer function with two Laguerre polynomials from appendix B of [22].

The total parallel current is then obtained from (64) using (30). For example, for $Z = 1$ the result is

$$j_\| = 1.64 \frac{f_t KcB}{\langle B^2 \rangle}(p_i' + p_e' - 0.74 n_e T_e' - 1.17 n_e T_i') + \frac{c(p_i' + p_e')}{B}\left[\left(1 - \frac{B^2}{\langle B^2 \rangle}\right) K + W\right]. \tag{65}$$

Due to the proportionality to $K$, the average parallel current $\langle j_\| B \rangle$ vanishes when the toroidal current is zero, as anticipated in [11,13]. Note however that the parallel current *before* averaging is still nonzero even in the $K \to 0$ limit due to $W$.

## 6. Discussion and conclusions

We have shown that perfectly quasi-isodynamic fields provide an important point of reference for understanding transport in optimized stellarator designs. For although quasi-isodynamic fields are generally far from quasisymmetry, neoclassical calculations are analytically tractable in a quasi-isodynamic field to the same extent as in a tokamak, with several modest modifications.

We have derived a general form for the flow in a quasi-isodynamic stellarator, given in (28)-(29). The form resembles the flow in an axisymmetric or quasisymmetric field, but a new term $W$ arises in the Pfirsch-Schlüter flow due to the deviation from symmetry. Our form of the flow agrees with the previously published distribution function for the long-mean-free-path regime, but the derivation here is valid also at higher collisionality. A similar form (30)-(31) for the parallel current immediately follows. The new quantity $W$ can be evaluated readily for a given $B(\theta, \zeta)$ using (43) or (45). When the enclosed toroidal current $cK/2$ vanishes in the long-mean-free-path regime, the parallel flow and current become proportional to $W$, and their flux surface averages vanish. For the more general case of nonzero $K$, we can use (45) to estimate

$$\frac{W}{K} \sim \frac{\left(\hat{B}-\breve{B}\right)}{B}\frac{I}{K} = \frac{\left(\hat{B}-\breve{B}\right)}{B}\frac{i_{\mathrm{p}}}{i_{\mathrm{t}}} \qquad (66)$$

where $i_{\mathrm{p}}$ is the poloidal current outside the flux surface (essentially equal to the total current in the external toroidal field coils) and $i_{\mathrm{t}}$ is the plasma current (the toroidal current inside the flux surface). As $i_{\mathrm{p}} \gg i_{\mathrm{t}}$ in any stellarator, the new $W$ terms in $V_{\parallel}$ and $j_{\parallel}$ will dominate over the "tokamak-like" terms proportional to $K$ even when $K$ is not strictly zero. For example, for W7-X parameters, even with a high estimate for the bootstrap current, $W/K > 60$. To a very good approximation then,

$$\mathbf{V}_a = \frac{c}{B^2}\left(\Phi' + \frac{p_a'}{Z_a e n_a}\right)\left(\mathbf{B}\times\nabla\psi + W\mathbf{B}\right) \qquad (67)$$

for each species $a$ and

$$\mathbf{j} = \frac{cp_{\mathrm{tot}}'}{B^2}\left(\mathbf{B}\times\nabla\psi + W\mathbf{B}\right). \qquad (68)$$

Due to the departure from symmetry, the radial particle flux is not intrinsically ambipolar. We can therefore solve for the radial electric field by imposing ambipolarity. Using a momentum-conserving pitch-angle-scattering model collision operator, the electric field in the long-mean-free-path regime is found to have the concise form (62). In the limit of quasi-poloidal symmetry, corresponding to $\left(\partial h / \partial \alpha\right)_B = 0$, the radial ion flux (54)-(57) becomes zero even when the velocity integral in (59) is not, so the electric field becomes undetermined.

These results for the flow, current, and radial electric field may be used to validate codes, since in a code it can be possible to specify a perfectly quasi-isodynamic field $B(\theta, \zeta)$. Also, in an optimized stellarator which is not perfectly quasi-isodynamic but nearly so, our results may apply approximately. In a field which is not perfectly quasi-isodynamic, a $D \propto 1/\nu$ regime will reappear for the radial particle flux, so the ambipolar radial electric field (62) will be the first of our results to break down. However, the parallel flows and current are less sensitive to the details of the ripple-trapped particles, so our results for the flows and current should be relatively robust.

**Acknowledgements**

This research was supported by the United States Department of Energy under grant DE-FG02-91ER-54109.

**Appendix A: Cary-Shasharina Theorem**

Consider the longitudinal invariant for trapped particles:

$$J(\psi, \alpha, \upsilon, \lambda) = \oint \upsilon_\| \, d\ell \tag{69}$$

where the integration is carried out along a full bounce. Using $d\ell = dB / \mathbf{b} \cdot \nabla B$ we can write

$$J = 2\upsilon \int_{\hat{B}}^{1/\lambda} dB \sqrt{1-\lambda B} \sum_\gamma \frac{\gamma}{\mathbf{b} \cdot \nabla B} \tag{70}$$

where the 2 arises from a sum over $\text{sgn}(\upsilon_\|) = \pm 1$, and the integral is performed at fixed $\psi$, $\alpha$, $\lambda$, and $\upsilon$. Applying a $(\partial / \partial \alpha)_{\psi,\upsilon,\lambda}$ derivative, and noting that $\partial J / \partial \alpha = 0$ due to omnigeneity, then

$$0 = \int_{\hat{B}}^{x} dB \, g_{\text{CS}} \sqrt{x-B} \tag{71}$$

where

$$g_{\text{CS}}(\psi, \alpha, B) = \left(\frac{\partial}{\partial \alpha}\right)_B \sum_\gamma \frac{\gamma}{\mathbf{b} \cdot \nabla B} \tag{72}$$

and $x = \lambda^{-1}$. The original definition of the longitudinal invariant (69) was valid for all trapped particles (i.e. for $\lambda > \hat{B}^{-1}$ and $\lambda < \check{B}^{-1}$), and so (71) is true for any $x$ between $\check{B}$ and $\hat{B}$. The proof in the appendix of [8] then applies, and so $g_{\text{CS}}$ must vanish at all locations.

**Appendix B: Construction of quasi-isodynamic fields**

Here we review the procedure described in [8] for constructing a field $B(\theta, \zeta)$ with the property (19) and with $\hat{B}$ straight in Boozer coordinates. While the original construction in [8] was given for a field in which the $B$ contours may close toroidally or helically, here we specialize to the case where the contours close poloidally. We have also modified the notation slightly to account for the possibility of multiple toroidal periods. As shown in section 2, any field generated by the construction which follows will be quasi-isodynamic.

We assume the stellarator has $N$ identical toroidal periods, with the $B = \hat{B}$ curves falling along $\zeta = 2\pi N^{(0)}/N$ for any integer $N^{(0)}$. We also assume there is a single $B = \breve{B}$ curve in each of the $N$ periods, and there are no local maxima or minima in $B$ aside from the global extrema $\hat{B}$ and $\breve{B}$. We define $\varepsilon_r = (\hat{B} - \breve{B})/(2\breve{B})$ and define $\eta$ by the relation
$$B/\breve{B} = 1 + \varepsilon_r + \varepsilon_r \cos\eta, \tag{73}$$
so contours of $B(\theta, \zeta)$ are contours of $\eta(\theta, \zeta)$. We stipulate that $\eta = 0$ at $\zeta = 0$, varying continuously to $\eta = 2\pi$ at $\zeta = 2\pi/N$. Our goal will be to construct $\eta(\theta, \zeta)$ for the single period $0 \leq \zeta \leq 2\pi/N$. We define
$$G_{\text{CS}} = N\zeta - \eta. \tag{74}$$
For $\eta$ in the range $0 \leq \eta \leq \pi$, $G_{\text{CS}}$ is then specified to be some continuous function of $\theta$ and $\eta$ such that $G_{\text{CS}} = 0$ when $\eta = 0$. For example, the model field shown in the figures was obtained using $G_{\text{CS}}(\theta, \eta) = [0.6 + 0.9\sin\theta]\sin(\eta/2)$. A function $\Delta\zeta(\eta)$ is also chosen such that $\Delta\zeta(0) = 2\pi/N$ and $\Delta\zeta(\pi) = 0$, and it must satisfy the periodicity condition $\Delta\zeta(\eta) = \Delta\zeta(2\pi - \eta)$. For the figures we choose the inverted tent function $\Delta\zeta = 2|\eta - \pi|/N$. While we chose $G_{\text{CS}}(\theta, \eta)$ freely in the range $0 \leq \eta \leq \pi$, in the range $\pi < \eta \leq 2\pi$, $G_{\text{CS}}(\theta, \eta)$ is fixed by the requirement (19). To derive the mathematical constraint which is placed on $G_{\text{CS}}$ by this requirement, consider points *x* and *y* in figure 1, two points on the same field line and at the same value of $B$ but on opposite sides of $\breve{B}$. Suppose point *x* has coordinates $\zeta = \zeta_0$, $\theta = \theta_0$, and $\eta = \eta_0$. Then point *y* has coordinates $\zeta = \zeta_0 - \Delta\zeta(\eta_0)$, $\theta = \theta_0 - \Delta\zeta(\eta_0)/q$, and $\eta = 2\pi - \eta_0$. Writing out (74) for each point, algebraically eliminating $\zeta_0$, and dropping the subscripts, we obtain
$$G_{\text{CS}}(\theta, \eta) = 2\pi - 2\eta + G_{\text{CS}}(\theta - \Delta\zeta(\eta)/q, 2\pi - \eta) + N\Delta\zeta(\eta). \tag{75}$$
This formula, which determines $G_{\text{CS}}$ for $\pi < \eta \leq 2\pi$ in terms of the $G_{\text{CS}}$ we chose for $0 \leq \eta \leq \pi$, ensures (19) is satisfied. Next, the relationship $\zeta(\theta, \eta) = [\eta + G_{\text{CS}}(\theta, \eta)]/N$ is inverted numerically to obtain $\eta(\theta, \zeta)$, and finally $B(\theta, \zeta)$ can be computed from (73).

For the model field shown in the figures, we have chosen $\varepsilon_r = 0.15$, $q = 1.079$, and $N = 5$.

**References**

[1]  Helander P and Sigmar D 2002 *Collisional Transport in Magnetized Plasmas* Cambridge University Press)

[2]  Hall LS and McNamara B 1975 *Phys. Fluids* **18** 552

[3]  Boozer AH 1995 *Plasma Phys. Control. Fusion* **37** A103

[4]  Nührenberg J and Zille R 1988 *Phys. Lett. A* **129** 113

[5]  Boozer AH 1982 *Phys. Fluids* **25** 520

[6]  Garren DA and Boozer AH 1991 *Phys. Fluids B* **3** 2822

[7]  Anderson F, Almagri A, Anderson D, Matthews P, Talmadge J and Shohet J 1995 *Fusion Tech.* **27** 273

[8]  Cary JR and Shasharina S 1997 *Phys. Plasmas* **4** 3323

[9]  Cary JR and Shasharina S 1997 *Phys. Rev. Lett.* **78** 674

[10] Gori S, Lotz W and Nührenberg J 1996 *Theory of Fusion Plasmas (Bologna: Editrice Compositori)* 335

[11] Subbotin AA et al. 2006 *Nucl. Fusion* **46** 921

[12] Mikhailov MI, Shafranov VD and Nührenberg J 2009 *Plasma Phys. Rep.* **35** 529

[13] Helander P and Nührenberg J 2009 *Plasma Phys. Control. Fusion* **51** 055004

[14] Lotz W, Nührenberg J and Schwab C 1991 *Proc. 13th Int. Conf. on Plasma Physics and Controlled Nuclear Fusion Research (Washington), Paper IAEA-CN-53-C-III-5, IAEA, Vienna*

[15] Greiger G et al. 1992 *Phys. Fluids B* **4** 2081

[16] Hirshman SP, Shaing KC, van Rij WI, Beasley, Jr. CO and Crume, Jr. EC 1986 *Phys. Fluids* **29** 2951

[17] Ho DD and Kulsrud RM 1987 *Phys. Fluids* **30** 442

[18] Wakatani M 1998 *Stellarator and Heliotron Devices* Oxford University Press)

[19] Helander P 2007 *Phys. Plasmas* **14** 104501

[20]  Simakov AN and Helander P 2009 *Phys. Plasmas* **16** 042503

[21]  Helander P and Simakov A 2008 *Phys. Rev. Lett.* **101** 145003

[22]  Pusztai I and Catto PJ 2010 *Plasma Phys. Control. Fusion* **52** 075016

**Figure 1.**

(Colour online.) Contours of $B$ for the model quasi-isodynamic field specified in appendix B. The dimensionless magnitude $b = B/\langle B^2 \rangle^{1/2}$ varies from 0.88 in the center to 1.15 at the left and right edges.

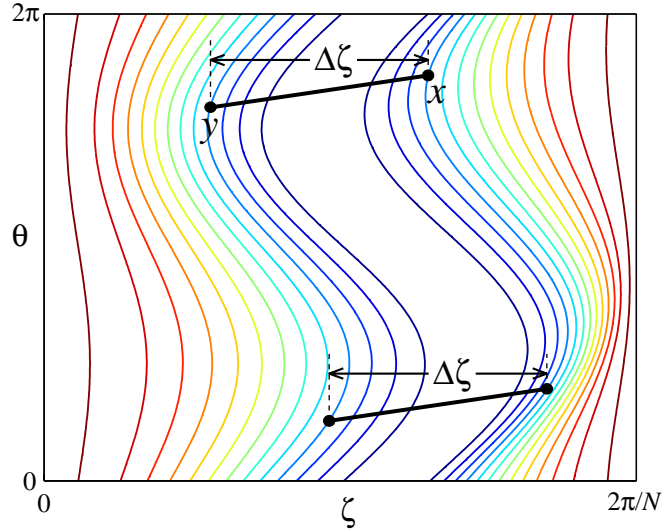

**Figure 2.**

(Colour online.) The departure from quasisymmetry $(\partial h/\partial\alpha)_B$, calculated for the model field of figure 1 using (16). Dashed curves are negative. The normalized quantity $q(qI+K)^{-1}(\partial h/\partial\alpha)_B$ ranges from -0.15 to 0.22.

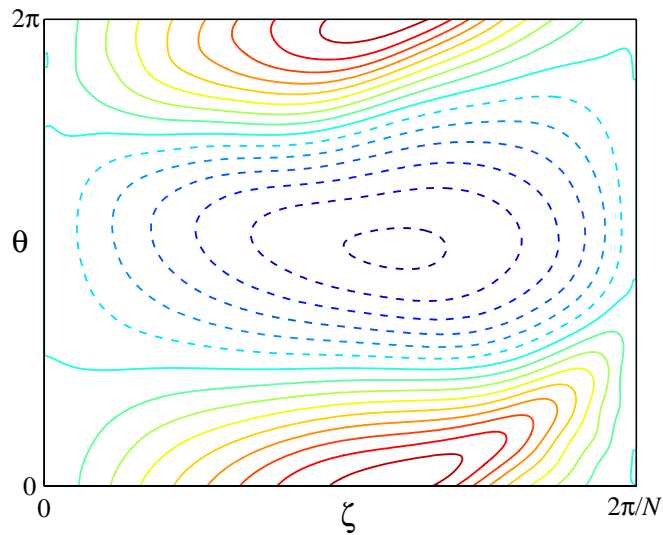

**Figure 3.**

(Colour online.) The integral $W$, defined by (25), which appears in the parallel flow and current. The calculation is performed for the model field of figure 1 using (43). Contours from -0.05 to 0.06 of the dimensionless quantity $Wq/(qI+K)$ are plotted. Dashed curves are negative.

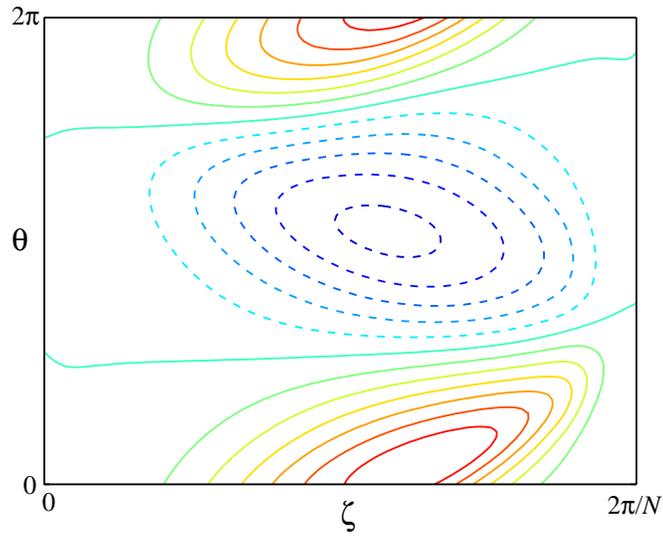